# Ultra-compact beam steering nanolasers

**Authors:** Xinghong Chen[1,2†], Mingxuan Gu[1†], Jiankai Tang[1†], Yungang Sang[3], Bingrui Xiang[1], Kong Zhang[1], Guanjie Zhang[1], Xingyuan Wang[4], Xuhan Guo[5], Linjie Zhou[6], Wengang Wu[7,8,9], Yifei Mao[1,2]*

**Affiliations:**

[1]School of Sensing Science and Engineering, School of Electronic Information and Electrical Engineering, Shanghai Jiao Tong University, Shanghai 200240, China.

[2]SJTU-Pinghu Institute of Intelligent Optoelectronics, Pinghu 314200, China.

[3]Institute of Microelectronics, Chinese Academy of Sciences, Beijing 100029, China

[4]College of Mathematics and Physics, Beijing University of Chemical Technology, Beijing 100029, China

[5]State Key Laboratory of Advanced Optical Communication Systems and Networks, Department of Electronic Engineering, Shanghai Jiao Tong University, Shanghai 200240, China

[6]State Key Laboratory of Advanced Optical Communication Systems and Networks, Shanghai Key Lab of Navigation and Location Services, Shanghai Institute for Advanced Communication and Data Science, Department of Electronic Engineering Shanghai Jiao Tong University, Shanghai 200240, China

[7]National Key Laboratory of Science and Technology on Micro/Nano Fabrication, School of Integrated Circuits. Peking University, Beijing 100871, China

[8]Beijing Advanced Innovation Center for Integrated Circuits, Beijing 100871, China

[9]Frontiers Science Center for Nano-optoelectronics, Peking University, Beijing 100871, China

*Correspondence to: maoyifei@sjtu.edu.cn

†These authors contributed equally to this work.

## Abstract

The miniaturization and integration of beam steering devices have consistently been the focus of the field. Conventional methods alter the eigenmode of the optical cavity by regulating the refractive index. Due to the weak nonlinear effect of the optical system, the device must be sufficiently large to achieve sufficient light modulation. The effective method for miniaturizing beam steering devices currently in use is based on metasurfaces. However, this type of device necessitates the input of a laser source, which precludes the simultaneous generation and control of light in a single device. Here we propose a miniaturized beam steering device that employs mode selection between different bound states in the continuum (BIC) states through phase change material. The device is capable of simultaneously achieving both light generation and beam steering (33°) in a single device with a size of only 25 μm and with a low threshold of 8.9 kW cm$^{-2}$. Furthermore, it is possible to achieve a significant degree of dynamic wavelength tunability, with a range extending up to 296 nm. This method achieves high-efficient regulation of light properties by dynamically controlling the system's topological charge, circumventing the problem of weak nonlinearity in traditional methods. Furthermore, the integration of phase change materials with nanolasers enables the direct alteration of lasing properties, which provides a novel idea for dynamic light control. The device process scheme based on phase change materials is straightforward, direct, and highly compatible, which will be advantageous for its intended application.

# Introduction

Optical beam steering, which involves the dynamic and precise control of light beams to specific directions, demonstrates pivotal applications across diverse domains, including as optical communications[1], light detection and ranging (LiDAR)[2-4] and imaging[3,5]. One of the current focuses is the miniaturization of beam steering devices. The reduction in size of a device can improve its performance, including response speed[6], power consumption[7], and integration[8]. Nevertheless, it is challenging to attain precise light control in small-scale devices. In conventional optical devices such as phase arrays[9,10] or modulators[2,11], the light-matter interaction is typically quite weak. The weak nonlinearity in an optical system implies that the alteration of optical parameters (e.g., refractive index) in response to external excitation will be relatively small. To achieve sufficient light field modulation, a device with a large size is typically required.

At present, the most effective approach to the miniaturization of beam steering devices is based on planar optical devices, such as metasurfaces[12-14]. Metasurfaces can manipulate the light field by introducing abrupt phase changes. They can be actively controlled by external stimuli, such as mechanical and chemical methods[15,16], to dynamically modulate the properties of incident light beams. Consequently, efficient beam steering can be accomplished at the nanoscale. However, the implementation of an active metasurface necessitates the combination of two essential components: an external light source and a beam control unit. These discrete components necessitate the use of disparate device technologies and material platforms, which are detrimental to future on-chip integration applications. The necessity for a compact beam steering device that is capable of simultaneously generating and regulating light at the nanoscale is currently of paramount importance.

Here, we propose a compact and active beam steering device based on bound states in the continuum (BIC), where laser generation and control can be performed simultaneously. We design different BICs at the Γ and off-Γ points in the momentum space. The BIC modes with varying momentum will be selected during lasing, resulting in distinct emission angles. Phase change materials are employed to regulate the mode competition between the two BICs. The schematic diagram of the device is shown in Figure 1a, where we integrate photonic crystals with a low-loss phase change material ($Sb_2Se_3$). In different crystal states of $Sb_2Se_3$, the device state can switch between $\Gamma$ BIC and off-$\Gamma$ BIC, thereby achieving an angle scanning of 33°. This work has two distinct advantages. On the one hand, the target BICs can be precisely selected during lasing to significantly alter the radiation performance. It can circumvent the issue of weak nonlinearity that is inherent to traditional methods. Consequently, the beam steering device has a record minimum size of 25 μm. In addition to the angle change, the device exhibits wavelength tunability of up to 296 nm, which is the largest among all tunable lasers at the near-infrared regime. On the other hand, the integration of phase change materials with nanolasers enables the direct alteration of laser performance. It allows for the generation and manipulation of laser beams to be accomplished simultaneously in a single compact device. Moreover, the introduction of low-loss phase change material result in a device threshold of 8.9 kW cm$^{-2}$ and 54.5 kW cm$^{-2}$ under two states of $Sb_2Se_3$, which is comparable to that of current low-threshold nanolasers.

## Beam steering nanolasers based on different BICs

The beam steering device is comprised of a multi-layered two-dimensional photonic crystal, as illustrated in Figure 1a. The lattice constant, $a$, and the radius of the hole, $r$, are 780 nm and 185 nm, respectively. The bottom layer is the 200 nm gain media, which is composed of an InGaAsP multilayer quantum well. A 10 nm $SiO_2$ layer is situated between gain media and phase change material to mitigate the effect of material loss on the lasing threshold. A low-loss $Sb_2Se_3$ layer with a thickness of 40 nm is covered in the photonic crystal for regulating the lasing properties. The top layer is 20 nm $SiO_2$, which serves to prevent the oxidation. As depicted in the SEM of Figure 1c, the dimensions of the beam steering device are approximately 25×25 μm. The inset shows a magnified detailed view of the cell array.

The BICs are polarization singularities in momentum space, with a theoretically infinite lifetime. The radiation characteristics of different BICs are distinct, including frequency, momentum, spin, and so forth. To achieve beam steering, two BICs are constructed. The first is a symmetric protective BIC at the Γ point (Γ BIC), while the second is an accidental BIC located far away from the Γ point (off-Γ BIC). The two BICs exhibit disparate momentum, which implies that they will have different emission angles during the lasing process. We simulate the band structure as shown in Figure 1b. Both BICs are TE-like modes and their electric field distribution of x-y plane are shown in the inset. The off-Γ BIC is in momentum space at $k_x a/(2\pi) = 0.264$, which means the theoretical emission angles of off-Γ BIC is 31°. The wavelength of the Γ and off-Γ BIC are 1423 nm and 1586 nm, respectively. BIC modes are far-field polarized topological defects carrying an integer topological charge $q^{17}$, defined as $q = \frac{1}{2\pi}\oint_C \nabla_k \varphi(\bm{k}) \cdot d\bm{k}$. Here C is a closed path in momentum space in the counterclockwise direction around the polarization singularity. $\varphi(\bm{k})$ is the angle of the polarization vector described by the angle between the major axis of polarization and the x component of the wave vector ($k_x$) and k is the in-plane wavevector. The topological charges of the two BIC modes are both +1 as shown in the inset. Once two BICs have been designed, the next objective is to select the desired BIC mode during lasing.

Different BICs are selected by regulating the states of the phase change material. We characterize the lasing characteristics of device in both amorphous $Sb_2Se_3$ (a-$Sb_2Se_3$) and crystalline $Sb_2Se_3$ (c-$Sb_2Se_3$), respectively. First, the device's performance is measured in the a-$Sb_2Se_3$ state. Figure 2a shows a single-mode lasing spectrum with a wavelength of 1545 nm. In the inset of Figure 2a is the K-space pattern, which illustrates the salient characteristics of off-Γ BIC. Given that the BIC mode is unable to be coupled to the far field, dark spots are observed in the center of the far field pattern. We further test the polarization of the far field pattern, indicating that the off-Γ BIC have a topological charge of +1 (Figure 2c). Figure 2e illustrates angle-resolved intensity profiles derived from far-field patterns, which indicate that the emission angle of off-Γ BIC is ~33°. The threshold of the laser can be extracted from the light-light curve (Figure 2g), and is approximately 8.9 kW cm$^{-2}$.

Subsequently, the sample is placed in a nitrogen environment for rapid annealing, which facilitates a state transformation of Sb$_2$Se$_3$ from its amorphous to crystalline state. The annealing process is initiated at a temperature of 180°C for 5 minutes, after which the sample is allowed to cool to room temperature. We then measure the corresponding lasing properties in the c-Sb$_2$Se$_3$ state (Figure 2b). In contrast to the device in the a-Sb$_2$Se$_3$ state, which exhibits only off-Γ BIC lasing, the device now exhibits lasing dominated only by Γ BIC. The wavelength of the single-mode lasing is 1419 nm. The results of the far-field polarization analysis (Figure 2d) indicate that the Γ BIC possesses a topological charge of +1. In the supplementary material S2, we test the self-interference pattern of the emission light. We find two obvious bifurcations of the interference fringes, indicating that the emission light of the Γ BIC mode is a vortex beam carrying topological charge of +1. Figure 2f illustrates angle-resolved intensity profiles of Γ BIC derived from far-field patterns, which indicates that the lasing direction is vertical. The measured lasing threshold is approximately 54.5 kW cm$^{-2}$ (Figure 2h). All the test results presented in Figure 2 demonstrate a general concordance with the simulation results displayed in Figure 1b. We also simulate the far-field patterns of the two BICs (Supplementary material S1), which is consistent with the test results.

The preceding results indicate that the switching between the two states of Sb$_2$Se$_3$ will result in a significant transition in the radiation properties of the device. Due to the different characteristics of different BICs, the emission angle of the beam steering device has undergone a notable change of 33°, while the lasing wavelength has exhibited a significant shift of 126 nm. The supplementary material S3 presents the radiation properties of devices with different lattice constants and hole diameters, which demonstrates a stable trend and regularity. The next question is to identify the factors that influence the selection between the two types of BICs during the lasing process.

**Mode competition during the phase transition of Sb$_2$Se$_3$**

The next paragraphs will discuss how the competition between different BICs alters due to the phase transition of $Sb_2Se_3$. Firstly, an experimental investigation is conducted to observe the competition between the $\Gamma$ BIC and off-$\Gamma$ BIC during the lasing process, as the state of $Sb_2Se_3$ gradually changes. Four gradient annealing temperatures is established to facilitate a gradual adjustment of the phase transition. Secondly, the causes of the mode competition are discussed. It is related to the variation of BICs' Q value with the material loss, as well as the relationship between the mode distribution and the gain spectrum.

Figure 3 illustrates the lasing properties when $Sb_2Se_3$ is in four distinct states. The red and blue curves represent the L-L curves of off-$\Gamma$ and $\Gamma$ BIC. Figure 3a shows the lasing spectrum in the a-$Sb_2Se_3$ state. The results indicate that only off-$\Gamma$ BIC lasing occurs, with a threshold pump power of 8.9 kW cm$^{-2}$. In the a-$Sb_2Se_3$ state, the off-$\Gamma$ BIC mode completely dominates the competition during the lasing process. Subsequently, the sample is subjected to annealing at a temperature of 160°C for 5 minutes (Figure 3b). During this period, $Sb_2Se_3$ is in the initial stages of a phase transition, yet has not yet undergone a complete change in phase. When the pump power increases to 31.4 kW cm$^{-2}$, off-$\Gamma$ BIC lasing mode appears first. Subsequently, when the pump power increases to 46.8 kW cm$^{-2}$, the sample starts to exhibit multi-mode lasing properties. The inset shows the K-space pattern at a pump power of 56.3 kW cm$^{-2}$, which indicates the coexistence of the $\Gamma$ BIC and off-$\Gamma$ BIC lasing modes. The threshold of the off-$\Gamma$ BIC is lower than that of the $\Gamma$ BIC mode. Next, the sample is annealed at 170°C for 5 minutes (Figure 3c) and $Sb_2Se_3$ continues to undergo phase transition. In this state, the device also exhibits multi-mode lasing. Different from Figure 3b, now the $\Gamma$ BIC mode appear first as the pump power increases. When the pump power increases to 67 kW cm$^{-2}$, the lasing state will produce a sudden change and that output energy of $\Gamma$ BIC will be much greater than that of off-$\Gamma$ BIC. The same phenomenon can be found in Figure 3b (when the power increases to 73.2 kW cm$^{-2}$). This is because the $\Gamma$ BIC at 1419 nm will receive more gain than off-$\Gamma$ BIC at 1545 nm as the pump power become larger. Finally, we anneal the sample at 180°C for 5 minutes to allow the $Sb_2Se_3$ fully transform into c-$Sb_2Se_3$ state. Regardless of the power increases, we can only observe

Γ BIC mode lasing. This means Γ BIC mode completely dominates the competition during the lasing process when the device is in the c-$Sb_2Se_3$ state.

The mode competition is a complex nonlinear process, making it challenging to engage in a quantitative explanation. Here we endeavor to conduct a brief discussion of the underlying reasons from two aspects. On the one hand, we analyze the relationship between the Q value of two BICs and the imaginary part of the $Sb_2Se_3$'s refractive index in different states. We know that when the device switches between different states, the primary change is the real and imaginary parts of the refractive index of $Sb_2Se_3$. In the supplementary material S4, we conduct a simulation to determine the Q factor of two BICs under different loss conditions. The Q factor of two BICs is highly sensitive to variations in the imaginary part of the refractive index. Consequently, the introduction of varying material losses in different phase states will alter the competitive relationship between Γ BIC and off-Γ BIC lasing. Nevertheless, as the simulation of the Q value involves the actual measured value of the extinction coefficient of $Sb_2Se_3$ (currently it is difficult to obtain accurate value). Therefore, we have made reasonable estimates but cannot draw firm conclusions (supplementary material S4). On the other hand, we investigate the relationship between different BIC mode distributions and gain spectrum. The selection of a lasing mode is also related to the material gain, which is influenced by the gain spectrum and the electromagnetic field distribution of modes within the gain material[18]. In the supplementary material S4, we present a simulation of the field distribution of the magnetic field Z component ($H_z$) for two BICs in the gain dielectric. It can be reasonably deduced that the off-Γ BIC is more dominant in the lasing process, as it interacts more effectively with the gain in the amorphous state. While in the crystalline state, an increase in pump power results in a notable enhancement of the gain spectrum intensity at 1419 nm (Γ BIC). This improvement enables the Γ BIC mode to achieve a greater gain, thereby facilitating its dominance.

## The impact of $Sb_2Se_3$ to the device performance

In this work, we introduce phase change materials into the photonic crystal nano-laser to produce beam steering. Phase change materials represent a mature approach of dynamic light field control. However, there have been few instances of their application to laser fields. This is because that a significant material loss will inevitably affect the lasing threshold, particularly in the optical and near-infrared regimes. Here, we use $Sb_2Se_3$, a novel phase change material exhibiting low material loss in either its amorphous or crystalline state. The following paragraphs will analyze the impact of the real and imaginary parts of $Sb_2Se_3$'s refractive index on the lasing properties. Then, the impact of material loss on the laser thresholds will be discussed.

Firstly, we analyze the effect of the real part of the refractive index on the lasing properties. Here, we alter the lattice constant and hole diameter to 760 nm and 510 nm, respectively. The thickness of the $Sb_2Se_3$ layer is set to 20 nm. Figure 4a illustrates the lasing spectrum before and after the phase transition. The gray and black lines represent the states of a-$Sb_2Se_3$ and c-$Sb_2Se_3$ respectively. The lasing wavelengths for a-$Sb_2Se_3$ and c-$Sb_2Se_3$ are 1541 nm and 1548 nm, respectively, with a wavelength shift of 7 nm. The K-space patterns of the two states are shown in the insets, which demonstrate that the lasing mode is also a $\Gamma$ BIC mode. The tuning of emission wavelength is a consequence of the change in the real part of the refractive index that occurs during the phase transition between the crystalline state and amorphous states.

In addition to its real part of refractive index, the imaginary part of $Sb_2Se_3$'s refractive index exerts a significant influence on the lasing properties. In the preceding paragraph, we demonstrated the utilization of this feature to facilitate the transition between $\Gamma$ BIC and off $\Gamma$ BIC, thereby enabling beam steering (Figure 2). Here, we demonstrate that large-scale wavelength tuning can also be realized by changing of the imaginary part of the refractive index. As shown in Figure 4b, the orange and green lines represent the lasing spectrum for a-$Sb_2Se_3$ and c-$Sb_2Se_3$, respectively, with a wavelength of 1614 nm and 1320 nm. A remarkable wavelength shift of 294 nm is attained. To the best of our knowledge, this represents the largest wavelength shift achieved in tunable lasers[19-25] in the near-infrared regime. The K-space patterns indicate that the two states are the $\Gamma$-BIC mode and a non-BIC trivial mode, respectively.

Conventional phase change materials, such as GeSbTe, inevitably suffer material losses in both their crystalline and amorphous states[26]. This has consistently posed a significant challenge to their optoelectronic applications, particularly in the laser fields. To circumvent this issue, we employ $Sb_2Se_3$, a phase-change material with low loss characteristics[27,28]. The four L-L curves depicted in Figure 4c represent four distinct lasing conditions. The red and blue curves represent the off-$\Gamma$ and $\Gamma$ BIC of the beam-steering device mentioned in Figure 2. The black curve represents the $\Gamma$ BIC in Figure 4a, when the thickness of c-$Sb_2Se_3$ is 20 nm. The green curve represents the trivial mode in Figure 4b. The log-scale curve depicted in the inset facilitates the observation of differences in lasing thresholds. A comparison of the black curve with the blue and green curves reveals that, in the crystalline state, the thicker the $Sb_2Se_3$ layer, the higher the threshold of the corresponding device. While a comparison of the red curve with the blue and green curves reveals that, regardless of the operational mode of the device, the threshold of device with a-$Sb_2Se_3$ is significantly lower than that of the device with c-$Sb_2Se_3$. The preceding observations permit the conclusion that the c-$Sb_2Se_3$ will result in greater material loss than a-$Sb_2Se_3$. This is consistent with our suspicion in the supplementary material S4 that the measured extinction coefficient of c-$Sb_2Se_3$ should not be equal to zero. The lasing thresholds of four conditions are 4.42 KW cm$^{-2}$ (red curve), 5.86 KW cm$^{-2}$ (black curve), 54.5 KW cm$^{-2}$ (blue curve) and 78.3 KW cm$^{-2}$ (green curve), which are relatively low in comparison to current low-threshold nanolasers. Supplementary materials S5 shows the spectra near the lasing threshold, proving that lasing occurred in all four cases.

## Discussion of device performance

Next, we discuss the advantages of our device and compare its performance with current works. Figure 5a shows all the miniaturized beam steering devices reported so far[6-8,29-34], where the horizontal axis is the device size and the vertical axis is the beam scanning angle. The devices may be classified into two main categories: One is discrete beam steering device (blue squares) requiring an external light source, such as the metasurface-based devices mentioned earlier. Another one is the integrated beam steering device (pink squares), wherein the generation and control of laser beams is accomplished concurrently. In addition to the beam steering device, we also compare our device with all the currently reported tunable nanolasers (blue circles)[19,20,22,23,25,35-45], as shown in Figure 5b. Tunable nanolasers can alter the eigenmode of a laser cavity in response to external excitation. They are regarded as a foundational technology for the development of miniaturized optoelectronic devices. All the detailed comparison information above can be found in the supplementary material S6.

The first advantage of our device is that it can be significantly reduced in size while maintaining satisfactory light control capabilities. Conventional methods employ nonlinear methods to regulate the refractive index to dynamically change the optical performance of the device. Nevertheless, the constraints of nonlinear effects render devices based on conventional methods to be considerable size. We present a novel light control method that enables the dynamic regulation of topological charges in optical systems, thus circumventing the challenges posed by nonlinear limitation. In comparison to the beam steering devices depicts in Figure 5a, our device exhibits the smallest size (25 μm). When compared with three integrated devices (pink squares)[8,30,31], our device is approximately one to two orders of magnitude smaller in size. Moreover, in comparison to all the tunable nanolasers in Figure 5b, our device is compact and can exhibit excellent dynamic wavelength tuning capabilities. The tuning wavelength range is 294 nm, which is better than most devices.

The second advantage of this device is that it is integrated, which means that the generation and regulation of light are completed in a single device. All the existing miniaturized beam steering devices employ passive micro-nano optical arrays or microcavities to alter the properties of the incident light. The necessity of an external light source represents a significant drawback for on-chip optoelectronic applications. In our device, phase change materials are employed to directly tuning the laser mode. This approach enables the generation and control of optical signals to be achieved simultaneously within a small device area. Therefore, in comparison to all other miniaturized metasurface-based devices depicts in the Figure 4a, our device is integrated, compact and has a satisfactory beam scanning range (33°).

Lastly, our device exhibits a low threshold, and its implementation is based on conventional IC processes and mature optoelectronic technologies. We employ a low-loss phase-change material, $Sb_2Se_3$, resulting in a device with an operating threshold of 8.9 kW cm$^{-2}$. This threshold is equivalent to that of existing low-threshold photonic crystal lasers[46,47] and will satisfy the loss requirements in practical applications. The fabrication of this device is entirely accomplished through conventional IC processes. Furthermore, phase-change material-based control technology is a well-established and commercially available technology. Thus, our solution is more application-oriented, in contrast to the technology of stress stretching, doping control, and microfluidics used in the tunable nanolasers.

## Conclusion

In conclusion, we present a compact and integrated beam steering device that can simultaneously generate and control light. Beam steering is achieved by mode selection between different BICs using phase change material. This method circumvents the issue of inadequate nonlinearity in traditional solutions, thus enabling the device to achieve the smallest scale (~25 μm), a notable beam angle change (~33°) and a low lasing threshold (~8.9 kW cm$^{-2}$). In addition, significant wavelength tuning, with a maximum at ~294 nm, can also be achieved. The fabrication process is straightforward and well-established, and can be directly compatible with existing optoelectronic technology. This dynamic control scheme can be applied to the regulation of other dimensions of light fields, including orbital angular momentum and chirality. This technology offers a novel and highly promising avenue for future applications in optical communications, LIDAR, and imaging.

## Methods

### Device fabrication

To fabricate ultra-compact beam steering laser, we use a nanostructured membrane of InGaAsP multiple quantum wells as gain media. The membrane is made up of 6 well layers embedded within barrier layers, topped by 10 nm of InP. The well layers consist of $In_{x=0.56}Ga_{1-x}As_{y=0.938}P_{1-y}$ with a thickness of 10 nm. The barrier layers are composed of $In_{x=0.734}Ga_{1-x}As_{y=0.57}P_{1-y}$ with a thickness of 20 nm. A 200 nm layer of $SiO_2$ is applied using plasma enhanced chemical vapor deposition to act as a dry etch hard mask. Electron beam lithography (EBL) is then used to define high-resolution nanoholes in a square lattice on the resist. Afterward, the $SiO_2$ hard mask is shaped using inductively coupled plasma (ICP) etching. Another ICP dry etching process is employed to etch through the 200 nm multi-quantum well layer once the resist is removed. The $SiO_2$ mask is then removed using an HF solution. To create a suspended membrane, the InP substrate is etched away with an $HCl:H_2O$ (3:1) solution. Next, we deposit a 10 nm transition layer of $SiO_2$ using inductively coupled plasma chemical vapor deposition equipment. Above the transition layer, we deposit 40 nm of $Sb_2Se_3$ with magnetron sputtering. Finally, we deposit 20 nm $SiO_2$ as an anti-oxidation layer with inductively coupled plasm[10]a chemical vapor deposition equipment. The refractive index and thicknesses of the $Sb_2Se_3$ were measured using a dual rotating-compensator Mueller matrix ellipsometer (ME-L ellipsometer, Wuhan Eoptics Technology Co., Wuhan, China). See supplementary material S7 for the complete fabrication process.

**Simulations**

Cavity modes are simulated by finite-element method. We use the periodic structure to obtain the band structure, quality factor, far field polarization and field distributions. The refractive indices of MQWs, $SiO_2$ and $Sb_2Se_3$ were set to 3.38, 1.46 and 3.4, respectively. A perfect matched layer is added in the simulation to serve as boundary condition in z direction. To simulate the Q value under the loss of the phase change material, we directly set the refractive index imaginary part of the material in the $Sb_2Se_3$ layer. The experimental observed modes are identified by comparing their eigenmode wavelengths and polarization in momentum space.

**Optical characterization**

The devices are pumped at room temperature by a pulsed laser at 1064 nm, where its pulse width and repetition rate are 5 ns and 12 kHz, respectively. To simplify the optical setup, an objective (100×, 0.9 numerical aperture) is used for the focus of excitation beam and collection of emission beam simultaneously. We used a diaphragm and lens to get a small pump spot. The signals are then guided to a near-infrared camera and a spectrometer for imaging and spectral analyzing. And a self-interference optical path is used to make the output beam interferes with itself. We verify whether the output beam is vortex beam by analyzing the interference pattern. See supplementary material S8 for the details of optical setup.

## Data availability

We declare that the data supporting the findings of this study are available within the paper.

## References

1  Tran, M. A. *et al.* Extending the spectrum of fully integrated photonics to submicrometre wavelengths. *Nature* **610**, 54-60 (2022).
2  Park, J. *et al.* All-solid-state spatial light modulator with independent phase and amplitude control for three-dimensional LiDAR applications. *Nature Nanotechnology* **16**, 69-76 (2020).
3  Kim, I. *et al.* Nanophotonics for light detection and ranging technology. *Nature Nanotechnology* **16**, 508-524 (2021).
4  Li, B., Lin, Q. & Li, M. Frequency–angular resolving LiDAR using chip-scale acousto-optic beam steering. *Nature* **620**, 316-322 (2023).
5  Gu, T., Kim, H. J., Rivero-Baleine, C. & Hu, J. Reconfigurable metasurfaces towards commercial success. *Nature Photonics* **17**, 48-58 (2022).
6  Wu, P. C. *et al.* Dynamic beam steering with all-dielectric electro-optic III–V multiple-quantum-well metasurfaces. *Nature Communications* **10**, 3654 (2019).
7  Nemati, A. *et al.* Controllable Polarization‐Insensitive and Large‐Angle Beam Switching with Phase‐Change Metasurfaces. *Advanced Optical Materials* **10**, 2101847 (2022).
8  Sakata, R. *et al.* Dually modulated photonic crystals enabling high-power high-beam-quality two-dimensional beam scanning lasers. *Nature Communications* **11**, 3487 (2020).
9  Sun, J., Timurdogan, E., Yaacobi, A., Hosseini, E. S. & Watts, M. R. Large-scale nanophotonic phased array. *Nature* **493**, 195-199 (2013).

10   Fu, X., Yang, F., Liu, C., Wu, X. & Cui, T. J. Terahertz Beam Steering Technologies: From Phased Arrays to Field‐Programmable Metasurfaces. *Advanced Optical Materials* **8**, 1900628 (2019).

11   Li, Q. *et al.* A Purcell-enabled monolayer semiconductor free-space optical modulator. *Nature Photonics* **17**, 897-903 (2023).

12   Hu, G. *et al.* Coherent steering of nonlinear chiral valley photons with a synthetic Au–WS2 metasurface. *Nature Photonics* **13**, 467-472 (2019).

13   Zhuang, X. *et al.* Active terahertz beam steering based on mechanical deformation of liquid crystal elastomer metasurface. *Light: Science & Applications* **12**, 14 (2023).

14   Berini, P. Optical Beam Steering Using Tunable Metasurfaces. *ACS Photonics* **9**, 2204-2218 (2022).

15   Arbabi, E. *et al.* MEMS-tunable dielectric metasurface lens. *Nature Communications* **9**, 812 (2018).

16   Schoen, D. T., Holsteen, A. L. & Brongersma, M. L. Probing the electrical switching of a memristive optical antenna by STEM EELS. *Nature Communications* **7**, 12162 (2016).

17   Zhen, B., Hsu, C. W., Lu, L., Stone, A. D. & Soljacic, M. Topological nature of optical bound states in the continuum. *Phys Rev Lett* **113**, 257401 (2014).

18   Pickering, T., Hamm, J. M., Page, A. F., Wuestner, S. & Hess, O. Cavity-free plasmonic nanolasing enabled by dispersionless stopped light. *Nature Communications* **5**, 4972 (2014).

19   Zhu, S., Shi, L., Xiao, B., Zhang, X. & Fan, X. All-Optical Tunable Microlaser Based on an Ultrahigh- *Q* Erbium-Doped Hybrid Microbottle Cavity. *ACS Photonics* **5**, 3794-3800 (2018).

20   Lu, T.-W., Wu, C.-C., Wang, C. & Lee, P.-T. Compressible 1D photonic crystal nanolasers with wide wavelength tuning. *Optics Letters* **42** (2017).

21   Freire-Fernández, F. *et al.* Magnetic on–off switching of a plasmonic laser. *Nature Photonics* **16**, 27-32 (2021).

22   Wang, Z. *et al.* On-chip tunable microdisk laser fabricated on Er3+-doped lithium niobate on insulator. *Optics Letters* **46**, 380-383 (2021).

23   Choi, J. H. *et al.* A high-resolution strain-gauge nanolaser. *Nat Commun* **7**, 11569 (2016).

24   Lu, T. W., Wang, C., Hsiao, C. F. & Lee, P. T. Tunable nanoblock lasers and stretching sensors. *Nanoscale* **8**, 16769-16775 (2016).

25   Deka, S. S. *et al.* Real-time dynamic wavelength tuning and intensity modulation of metal-clad nanolasers. *Optics Express* **28**, 27346-27357 (2020).

26   Abdollahramezani, S. *et al.* Electrically driven reprogrammable phase-change metasurface reaching 80% efficiency. *Nature Communications* **13**, 1696 (2022).

27   Delaney, M., Zeimpekis, I., Lawson, D., Hewak, D. W. & Muskens, O. L. A New Family of Ultralow Loss Reversible Phase‐Change Materials for Photonic Integrated Circuits: Sb2S3 and Sb2Se3. *Advanced Functional Materials* **30**, 2002447 (2020).

28   Yang, X. *et al.* Non‐Volatile Optical Switch Element Enabled by Low‐Loss

Phase Change Material. *Advanced Functional Materials* **33** (2023).

29  Hutchison, D. N. *et al.* High-resolution aliasing-free optical beam steering. *Optica* **3**, 887 (2016).

30  Kurosaka, Y. *et al.* On-chip beam-steering photonic-crystal lasers. *Nature Photonics* **4**, 447-450 (2010).

31  Cho, S., Yoshida, H. & Ozaki, M. Emission Direction‐Tunable Liquid Crystal Laser. *Advanced Optical Materials* **8**, 2000375 (2020).

32  Xie, Y.-Y. *et al.* Metasurface-integrated vertical cavity surface-emitting lasers for programmable directional lasing emissions. *Nature Nanotechnology* **15**, 125-130 (2020).

33  Yin, X. *et al.* Beam switching and bifocal zoom lensing using active plasmonic metasurfaces. *Light: Science & Applications* **6**, e17016-e17016 (2017).

34  Takeuchi, G. *et al.* Thermally controlled Si photonic crystal slow light waveguide beam steering device. *Optics Express* **26**, 11529-11537 (2018).

35  Jin, L. *et al.* Dual-wavelength switchable single-mode lasing from a lanthanide-doped resonator. *Nature Communications* **13**, 1727 (2022).

36  Zhuge, M.-H. *et al.* Fiber-Integrated Reversibly Wavelength-Tunable Nanowire Laser Based on Nanocavity Mode Coupling. *ACS Nano* **13**, 9965-9972 (2019).

37  Yang, A. *et al.* Real-time tunable lasing from plasmonic nanocavity arrays. *Nature Communications* **6**, 6939 (2015).

38  Liu, G. *et al.* Toward Microlasers with Artificial Structure Based on Single-Crystal Ultrathin Perovskite Films. *Nano Letters* **21**, 8650-8656 (2021).

39  Qiao, Z. *et al.* Tunable Optical Vortex from a Nanogroove-Structured Optofluidic Microlaser. *Nano Letters* **22**, 1425-1432 (2022).

40  Zapf, M. *et al.* Dynamical Tuning of Nanowire Lasing Spectra. *Nano Letters* **17**, 6637-6643 (2017).

41  Chen, W. *et al.* Self-Assembled and Wavelength-Tunable Quantum Dot Whispering-Gallery-Mode Lasers for Backlight Displays. *Nano Letters* **23**, 437-443 (2023).

42  Yang, X. *et al.* An Electrically Controlled Wavelength-Tunable Nanoribbon Laser. *ACS Nano* **14**, 3397-3404 (2020).

43  Lv, Y. *et al.* Steric-Hindrance-Controlled Laser Switch Based on Pure Metal–Organic Framework Microcrystals. *Journal of the American Chemical Society* **141**, 19959-19963 (2019).

44  Wang, Y. *et al.* Tunable whispering gallery modes lasing in dye-doped cholesteric liquid crystal microdroplets. *Applied Physics Letters* **109**, 231906 (2016).

45  Gao, Z. *et al.* Proton-Controlled Organic Microlaser Switch. *ACS Nano* **12**, 5734-5740 (2018).

46  Shao, Z.-K. *et al.* A high-performance topological bulk laser based on band-inversion-induced reflection. *Nature Nanotechnology* **15**, 67-72 (2019).

47  Sang, Y. G. *et al.* Topological polarization singular lasing with highly efficient radiation channel. *Nat Commun* **13**, 6485 (2022).

## Acknowledgements

This work was supported in part by the Open Project Program of SJTU Pinghu Institute of Intelligent Optoelectronics (No. 2022SP10E202), the Shanghai Jiao Tong University 2030 Initiative (No. WH510363001-13), and the Research Fund of Shanghai Jiao Tong University Explore X. We thank the AEMD center of SJTU for providing fabrication service. The authors sincerely thank Prof. Renmin Ma from PKU for discussion, Zengkai Shao, Jingyu Lu, Shaolei Wang, Ziwei Zhao for helpful discussion, and Huazhou Chen for his guidance on simulations. We thank Wei Peng from Wuhan Eoptics Technology Co. for discussion of refractive index measurements.

## Author Contributions

Y. F. Mao and X. H. Chen conceived and supervised the project. X. H. Chen and M. X. Gu conducted the design. X. H. Chen, M. X. Gu, J. K. Tang, B. R. Xiang and X. Y. Wang performed the numerical calculations. X. H. Chen, K. Zhang and G. J. Zhang fabricated the sample. M. X. Gu, J. K. Tang and Y. G. Sang measured the device characteristics. X. H. Chen, L. J. Zhou and X. H. Guo discussed and analyzed the material properties of $Sb_2Se_3$. Y. F. Mao, X. H. Chen, M. X. Gu, J. K. Tang and W. G. Wu wrote the paper with input from all authors.

## Competing interests

The authors declare no competing interests.

# Figures

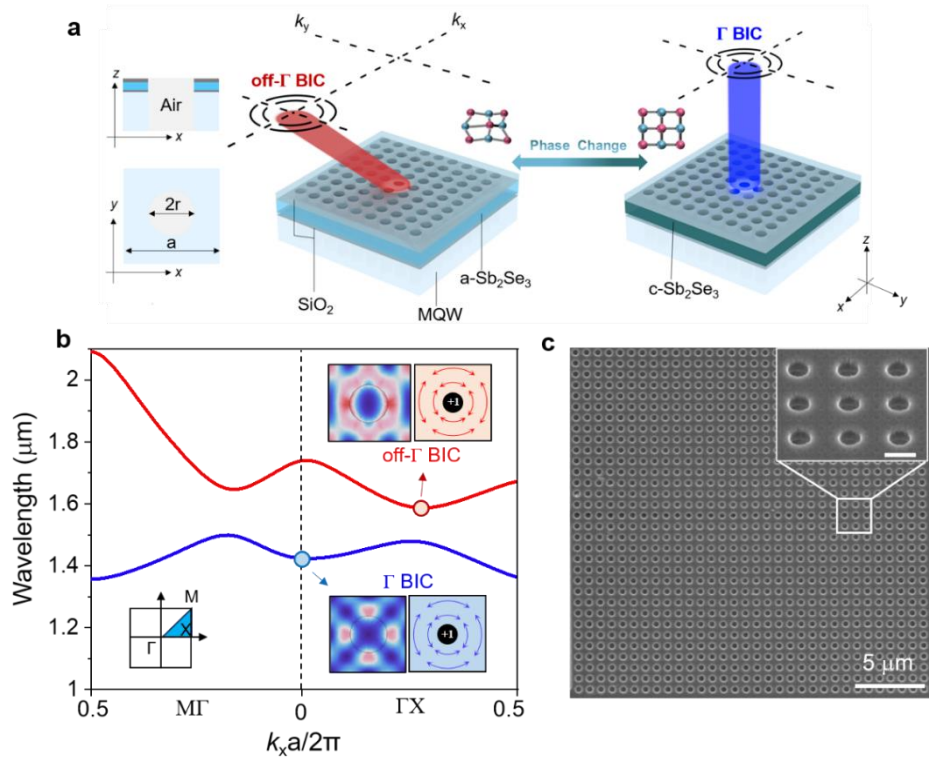

**Fig.1 A compact beam steering nanolaser. a,** Illustrations of the beam steering nanolaser. The radiation direction can be dynamically tuned during phase transition of $Sb_2Se_3$. Here a-$Sb_2Se_3$ and c-$Sb_2Se_3$ represent amorphous and crystalline $Sb_2Se_3$, respectively. $a$ is the lattice constant and $r$ is the air hole radius. **b,** Simulated band diagram of the device. The inserts depict the electric field distributions and far-field polarization of Γ BIC and off-Γ BIC, with both exhibiting a topological charge $q = +1$. The off-Γ BIC is located at $k_x a/(2\pi) = 0.264$, with the theoretical radiation angle of 31°. **c,** SEM image of a compact 25 μm×25 μm device. A semiconductor membrane serves as photonic crystal and gain material simultaneously and is covered with a 40 nm layer of $Sb_2Se_3$. The insert is a zoomed-in view, with a scale bar of 500 nm.

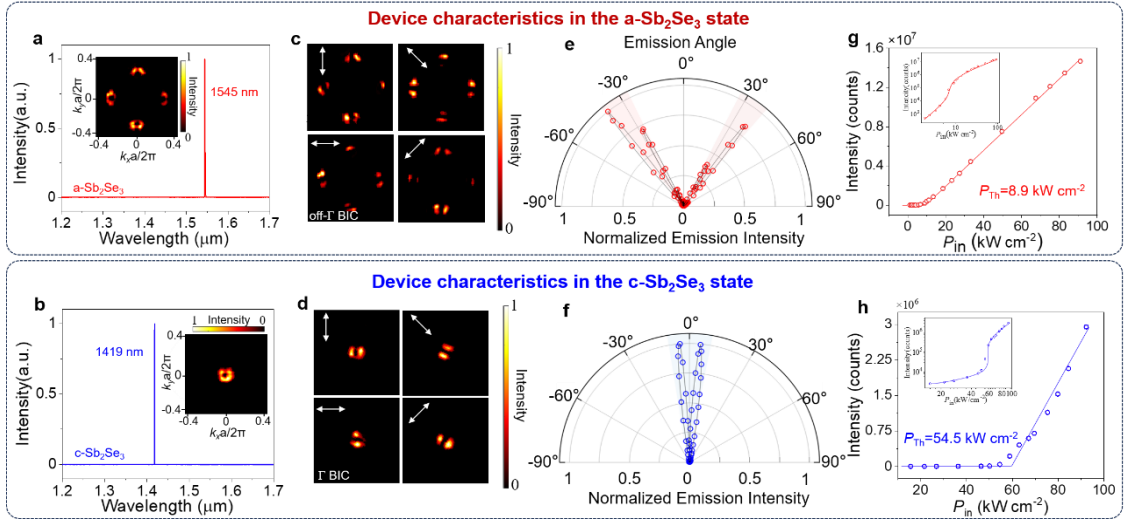

**Fig.2 Characterization of the beam steering nanolaser. a, b,** the lasing spectrum of device with a-$Sb_2Se_3$ (**a**) and c-$Sb_2Se_3$ (**b**). The insets are lasing emission pattern in momentum space, showing the characteristics of off-Γ BIC and Γ BIC, respectively. **c, d,** Lasing patterns in momentum space after a linear polarizer with varied polarization angles. Arrows: directions of the linear polarizer. **e, f,** Angle-resolved intensity profiles of off-Γ BIC (**e**) and Γ BIC (**f**). The variation in lasing angle reaches ~33°. Circles, data; lines, fitting. **g, h,** Integrated output intensity as a function of peak pump intensity on a linear scale for the device with a-$Sb_2Se_3$ (**g**) and c-$Sb_2Se_3$ (**h**). Insets: light-light curves on a logarithmic scale. Circles, data; lines, fitting. The lasing threshold of device with a-$Sb_2Se_3$ and c-$Sb_2Se_3$ are 8.9 kW cm$^{-2}$ and 54.5 kW cm$^{-2}$, respectively.

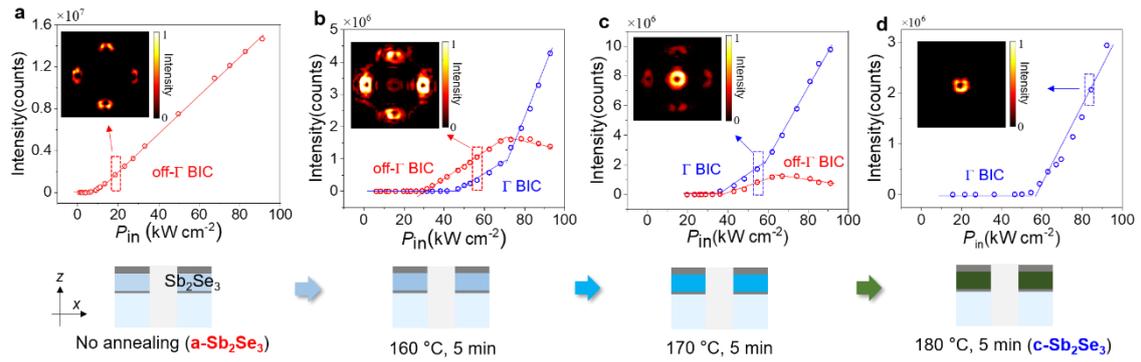

**Fig.3 BIC modes competition during phase transition. a-d,** Measured L-L curves and far-field emission patterns as the state of $Sb_2Se_3$ gradually changes. Four gradient annealing situations are set: no annealing (**a**), annealing at 160°C (**b**), annealing at 170°C (**c**) and annealing at 180°C (**d**). During the phase transition, an obvious mode competition process between the Γ BIC and off-Γ BIC are observed. When $Sb_2Se_3$ is in the amorphous state (**a**), only off-Γ BIC can be observed. As $Sb_2Se_3$ is in the intermediate phase transition state (**b, c**), both Γ and off-Γ BIC lasing appear. After $Sb_2Se_3$ is completely transformed into crystalline state (**d**), Γ BIC lasing completely dominates.

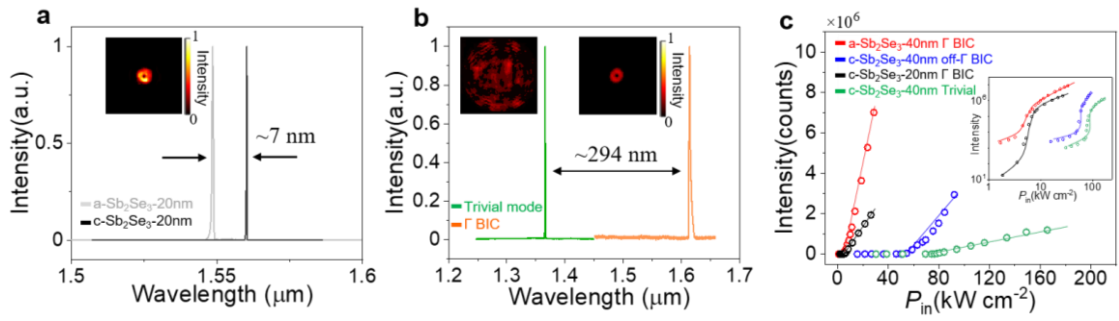

**Fig. 4 The impact of phase change material to the device performance. a,** Changes in the lasing wavelength when the real part of the refractive index of $Sb_2Se_3$ changes. The gray (black) lines represent the device with 20 nm a-$Sb_2Se_3$ (c-$Sb_2Se_3$). The wavelength variation is ~7 nm. The inset depicts the far-field pattern. **b,** Changes in the lasing wavelength when the imaginary part of the refractive index of $Sb_2Se_3$ changes. The orange (green) line represents a $\Gamma$ BIC (trivial) mode when the device is in the a-$Sb_2Se_3$ (c-$Sb_2Se_3$) state. The wavelength variation is ~294 nm. **c,** Measured L-L curves of four conditions, showing the effect of $Sb_2Se_3$ on lasing thresholds. The thresholds of the devices in the four conditions are 4.42 KW cm$^{-2}$ (red curve), 5.86 KW cm$^{-2}$ (black curve), 54.5 KW cm$^{-2}$ (blue curve) and 78.3 KW cm$^{-2}$ (green curve), respectively. Inset: curve on a log scale. Circles represent the data and curves represent the fitting.

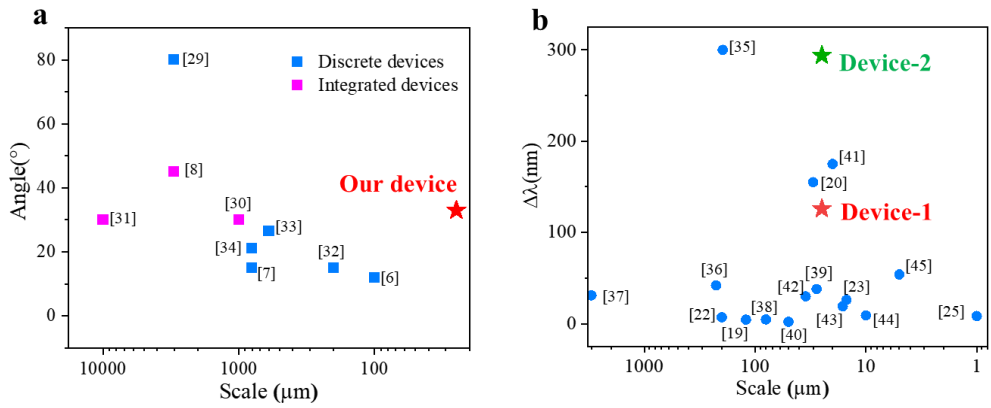

**Fig.5 Comparison of our device with current beam steering devices and tunable nanolasers. a,** Comparison of the device size and the scanning angle for currently miniaturized beam steering devices. The blue squares represent the discrete beam steering devices requiring external light sources. The pink squares represent the integrated devices, wherein the generation and control of laser beams is accomplished simultaneously. **b,** Comparison of device size and the wavelength variation of currently tunable nanolasers. "Device-1" refers to the device discussed in Figure 2, and "Device-2" refers to the device discussed in Figure 4b. The numbers in brackets in the figure represent reference numbers.